\documentclass[11pt,sort&compress]{elsarticle}
\usepackage[paper=a4paper,top=24mm, bottom=26mm, left=26mm, right=26mm]{geometry}
\makeatletter
\def\ps@pprintTitle{%
 \let\@oddhead\@empty
 \let\@evenhead\@empty
 \def\@oddfoot{\centerline{\thepage}}%
 \let\@evenfoot\@oddfoot}
\makeatother
\usepackage{natbib}
\usepackage[hyperref]{xcolor}
\definecolor{darkgreen}{rgb}{0.01, 0.75, 0.24}
\definecolor{darkblue}{HTML}{2B66D3}
\usepackage[colorlinks=true,
            linkcolor=darkblue,
            urlcolor=darkblue,
            citecolor=darkblue,linkcolor=darkblue,hyperfootnotes=true]{hyperref}
\usepackage{amsmath}
\usepackage{amsfonts}
\usepackage{slashed}
\usepackage{accents}
\usepackage{amssymb}
\usepackage{mathrsfs}
\usepackage{mathtools}
\usepackage[utf8]{inputenc}
\usepackage[T1]{fontenc}
\let\oldbibliography\thebibliography
\renewcommand{\thebibliography}[1]{%
  \oldbibliography{#1}%
  \setlength{\itemsep}{1.4pt}%
}

\usepackage[cal=boondox,calscaled=1]{mathalfa}
\DeclareMathAlphabet{\bbvar}{U}{BOONDOX-ds}{m}{n}

\usepackage{psfrag}
\usepackage{pstool}
\usepackage{caption}

\usepackage{tabularx}
\usepackage{multirow}
\usepackage{pbox}
\usepackage{graphicx}
\newcommand{\qq}[1]{``#1''} 
\newcommand{\utilde}[1]{\underaccent{\tilde}{#1}}
\newcommand{\di}{\mathrm{d}}
\usepackage{tensor}
\newcommand{\ou}[3]{\tensor{#1}{^{#2}_{#3}}}
\newcommand{\uo}[3]{\tensor{#1}{_{#2}^{#3}}}

\newcommand{\I}{\mathrm{i}} 
\newcommand{\C}{\mathbb{C}}

\newcommand{\R}{\mathbb{R}}

\newenvironment{subalign}{\subequations\align}{\endalign\endsubequations}
\newcommand{\eref}[1]{(\ref{#1})}

\DeclareMathAlphabet{\bbgreek}{U}{bbold}{m}{n}

\usepackage{tikz-cd}

\newcommand\vpm{\mathbin{\vcenter{\hbox{
  \oalign{\hfil$\scriptstyle+$\hfil\cr
          \noalign{\kern-.3ex}
          $\scriptscriptstyle({-})$\cr}}}}}
\DeclareMathAlphabet{\sfit}{OT1}{fos}{sb}{it}
\DeclareMathAlphabet{\mathsf}{OT1}{fos}{sb}{n}

\definecolor{darkgreen}{rgb}{0.01, 0.75, 0.24}
\definecolor{darkblue}{HTML}{2B66D3}

\usepackage[multiple, flushmargin]{footmisc}

\let\originalleft\left
\let\originalright\right
\renewcommand{\left}{\mathopen{}\mathclose\bgroup\originalleft}
\renewcommand{\right}{\aftergroup\egroup\originalright}

\newcommand{\dbarvar}{{\mathrm{d}\mkern-7.5mu\lower.18ex\hbox{$\textasciitilde$}\mkern-1.5mu}}

\renewcommand{\emph}[1]{{\it #1}}

\begin{document}

\begin{abstract}
In perturbative gravity, it is straight-forward to characterize the two local degrees of freedom of the gravitational field in terms of a mode expansion of the linearized perturbation. In the non-perturbative regime, we are in a more difficult position. It is not at all obvious how to construct Dirac observables that can separate the gauge orbits. Standard procedures rely on asymptotic boundary conditions or formal Taylor expansions of partial observables along the gauge orbits. In this paper, we lay out a new non-perturbative lattice approach to tackle the problem in terms of Ashtekar's self-dual formulation. Starting from a simplicial decomposition of space, we introduce a local kinematical phase space at the lattice sites. At each lattice site, we introduce a set of constraints that replace the generators of the hypersurface deformation algebra in the continuum. We show that the discretized constraints close under the Poisson bracket. The resulting reduced phase space describes two complex physical degrees of freedom representing the two radiative modes at the discretized level. The paper concludes with a discussion of the key open problems ahead and the implications for quantum gravity.
\end{abstract}
\title{Simplicial Graviton from Selfdual Ashtekar Variables}
\author{Wolfgang Wieland}
\address{Institute for Quantum Gravity, Theoretical Physics III, Department of Physics\\Friedrich-Alexander-Universität Erlangen-Nürnberg, Staudtstra\ss e 7, 91052 Erlangen, Germany\\{\vspace{0.5em}\normalfont03 May 2023}
}

\maketitle
{\vspace{-1.2em}
\hypersetup{
  linkcolor=black
}
{\tableofcontents}

\hypersetup{
  linkcolor=darkblue,
  urlcolor=darkblue,
  citecolor=darkblue
}
\begin{center}{\noindent\rule{\linewidth}{0.4pt}}\end{center}\newpage
\section{Introduction}\noindent

\noindent
Many non-perturbative approaches to quantum gravity start from discrete truncations, for example a lattice \cite{Loll:2019rdj,alexreview,rovelli,Surya:2019ndm,Ambjorn:2022naa,Dona:2022yyn}. When the lattice is finite, the truncation yields a finite-dimensional mechanical system that approximates the dynamics of general relativity in the continuum. In principle, such mechanical systems can be quantized at the full non-perturbative level.  A smooth semi-classical geometry, so it exists, can then only emerge after infinite refining \cite{Dittrich:2014ala,Steinhaus:2020lgb,Asante:2022dnj, Benedetti:2007pp,Ambjorn:2020rcn}. 
The main difficulties with this approach arise from the gauge symmetries of the theory in the continuum. 
In the continuum, the symmetries of Einstein's equations gives rise to a vast gauge redundancy.  The corresponding gauge generators are the scalar and vector constraints for the initial data. Gauge-invariant physical states are represented by gauge orbits on the constraint hypersurface. 
Although we have no explicit construction of such gauge-invariant physical states on a lattice,\footnote{Except in three dimensions, where there are no radiative modes.} there is a simple proposal for how to proceed to quantum theory. To characterize physical states, we would start from a discrete representation of the constraint algebra. Once we have such a representation, we turn to quantum theory to establish a suitable kinematical Hilbert space that carries a representation of the constraints. On this auxiliary Hilbert space, the constraints turn into the generators of gauge symmetries. The physical Hilbert space is the kernel of the now discretized constraints. A concrete realization of this idea can be found in the loop gravity and spinfoam approaches to quantum gravity \cite{rovelli, thiemann,qsd,LOSTtheorem, alexreview,Haggard:2023tnj}.  


One of the main difficulties with this idea is the issue of anomalies \cite{outside,Dittrich:2012qb}. It is not at all obvious how to put the constraints on a lattice without violating the gauge symmetries. If the gauge symmetry is broken, the constraints no longer form a closed algebra and we can no longer impose them strongly. The best that we could do is to impose them weakly, e.g.\ by restricting ourselves to a subspace of the Hilbert space, where all matrix elements of the constraints vanish. The fundamental problem with this idea is that there is no reason to believe that such a subspace would be stable under the dynamics. If it is not, transition amplitudes will no longer be unitary. Processes could occur where initial states that are sharply peaked on a classical configuration would decay into final states that violate gauge invariance strongly. 

Thus there seems to be a fundamental conflict between diffeomorphism invariance and lattice truncations. To resolve the problem, we could either resort to an infinite lattice refinement or work with an appropriately deformed constraint algebra. In the first case, diffeomorphism invariance is restored in the continuum limit \cite{Dittrich:2014ala,Asante:2022dnj}. In the second case, the lattice stays, but the gauge symmetries are deformed in such a way that there is no longer an anomaly. 
In this paper, we consider this second possibility \cite{qsd,Thiemann:2021hpa,Varadarajan:2022dgg,Tomlin:2012qz}.   By introducing a new discretization scheme, we obtain an anomaly-free lattice regularization for selfdual gravity \cite{newvariables,ashtekar}. Compared to the situation in the continuum, we have two additional constraints. Besides the Gauss, vector and Hamilton constraints, there is a new closure constraint and a cocylce condition for certain magnetic fluxes. Otherwise the algebra does not close. We will argue below that the additional constraints vanish in the continuum and are necessary to remove otherwise unphysical lattice artefacts. A simple counting demonstrates that the resulting reduced phase space has $2 \times 2$ complex dimensions per lattice site. This is an encouraging result, for it agrees with the situation in the continuum.  
In the continuum, self-dual gravity has two complex propagating modes. 

Our construction consists of two steps with two main results. The first result provides a new lattice regularization of the constraints on a fixed triangulation. The second, and more important result concerns the resulting constraint algebra, which closes under the Poisson bracket without any unwanted anomalies. At the same time, there are two important caveats. The first caveat is that we restrict ourselves in this paper to the classical level. At the quantum level, new anomalies may arise through operator ordering ambiguities. The second caveat is more severe. Our construction assumes that the constraints can be solved locally. The basic idea is to split the initial hypersurface into fundamental building blocks, then solve the discretized constraints in each building block separately. It is only in a second step that the fundamental cells  are  glued back together such that a new and more complicate solution of the constraint equations is found. This entire gluing procedure requires auxiliary boundary modes \cite{Balachandran:1994up,Geiller:2017xad,Wieland:2017zkf,Wieland:2017cmf,Donnelly:2016auv,Freidel:2023bnj} to regularize the constraints in each fundamental building block. In the present framework, these boundary modes are given by a flat boundary connection $A=g^{-1}\di g$ at the two-dimensional boundary of each cell on the three-dimensional Cauchy surface.\medskip

\emph{Outline of the paper.}  In \hyperref[sec2]{section 2}, we give a concise review of selfdual gravity as a constrained Hamiltonian system. 
The two main results are developed in \hyperref[sec3]{section 3} and \hyperref[sec4]{section 4}. In \hyperref[sec3]{section 3}, we introduce the quasi-local regularization of the constraints in a single building block. It is in this section that we also identify the additional closure constraints and cocycle conditions that remove otherwise unphysical lattice artefacts. The constraint algebra and corresponding structure functions are computed in \hyperref[sec4]{section 4}. The paper concludes with an outlook and discussion about the relevance of our results for non-perturbative approaches to quantum gravity such as loop quantum gravity and group field theory.\medskip

\emph{Notation.} In the following, indices $a,b,c,\dots$ are abstract (co)tangent indices on the spatial manifold $M$, 
greek indices $\alpha,\beta,\gamma,\dots=1,2,3$ refer to a fiducial coordinate system on $M$, indices $I,J,K,\dots=1,\dots,4$ (from the center of the alphabet) refer to the four sides of a tetrahedron  $T\subset M$ (see \hyperref[fig1]{figure 1}), indices $A,B,C,\dots$ are left-handed spinor indices, which are raised and lowered by the skew symmetric $\epsilon$-tensor\footnote{i.e.\ $\xi^A=\epsilon^{AB}\xi_B$, $\xi_A=\epsilon_{BA}\xi^B$.} and indices $i,j,k,\dots=1,2,3$ refer to a complexified basis in $\mathfrak{sl}(2,\C)$, which consists of the Pauli matrices that satisfy the familiar Pauli identity \eref{Pauliident}. With respect to this complexified basis, the structure constants of $\mathfrak{sl}(2,\C)$ are given by the components of the usual three-dimensional internal Levi-Civita tensor $\ou{\epsilon}{i}{jk}$. Finally, $\delta_{ij}$ denotes the corresponding internal three-metric, i.e.\ $\delta_{ij}=-\tfrac{1}{2}\ou{\epsilon}{l}{im}\ou{\epsilon}{m}{jl}$.
 

\section{Review: Phase space of selfdual gravity}\label{sec2}\noindent
Selfdual (complex) gravity admits a remarkably simple Hamiltonian formulation \cite{newvariables,ashtekar}. In the following, we briefly summarize the formalism. More recent developments on the subject can be found in \cite{Ashtekar:2020xll,Alexander:2022ocp,Eder:2020okh,Krasnov:2022mvn}.  This section will be  useful for the coherence of the presentation and the discussion of our results in the conclusion.\medskip

In selfdual gravity, the fundamental configuration variable is the self-dual and $\mathfrak{sl}(2,\C)$-valued Ashtekar connection $\ou{A}{i}{a}$ on the initial hypersurface. Its conjugate momentum is the densitized triad $\uo{\tilde{E}}{i}{a}$, which is an $\mathfrak{sl}(2,\C)$-valued vector density. At the kinematical level, the fundamental Poisson brackets are given by
\begin{equation}
\big\{\uo{\tilde{E}}{i}{a}(x),\ou{A}{j}{b}(y)\big\}=8\pi\I G \delta^j_i\delta^a_b\tilde\delta^{(3)}(x,y),\label{contpoiss}
\end{equation}
where $G$ is Newton's constant and $\tilde\delta^{(3)}(x,y)$ denotes the Dirac distribution (a scalar density of weight one). All other Poisson brackets among $\ou{A}{i}{a}$ and $\uo{\tilde{E}}{i}{a}$ vanish. Tensor indices $a,b,c,\dots$ are abstract tangent indices on the spatial hypersurface $M$. The internal indices $i,j,k,\dots=1,2,3$ refer to a three-dimensional basis in the $\mathfrak{sl}(2,\C)$  gauge fibres over every point of the manifold. This basis satisfies the usual Pauli identity
\begin{equation}
\ou{\tau}{A}{Ci}\ou{\tau}{C}{Bj}=-\frac{1}{4}\delta_{ij}\delta^A_B+\frac{1}{2}\uo{\epsilon}{ij}{k}\ou{\tau}{A}{Bk},\label{Pauliident}
\end{equation}
where $A,B,C,\dots$ are spinor indices for two-component Weyl spinors. The basis is complex, which is to say that $\mathfrak{sl}(2,\C)$ is treated as the complexification of $\mathfrak{su}(2)$. Real (hermitian) and imaginary (anti-hermitian) parts are computed with respect to a hermitian metric $\delta_{\bar{A}A}$. Given an embedding of the initial hypersurface $M$ into the four-dimensional manifold, this metric can be uniquely fixed as follows. If $\bar{\sigma}_{\bar{A}A\alpha}$ are the four-dimensional Infeld--Van der Waerden symbols, and $n^a$ is the time-like normal vector to $M$, the metric in the spin bundle is given by $\delta_{\bar A A}=\bar{\sigma}_{\bar AA\alpha}\ou{e}{\alpha}{a}n^a$, where $\ou{e}{\alpha}{a}$ is the co-tetrad. It is with respect to this metric that the Pauli matrices \eref{Pauliident} are anti-hermitian, i.e.\
\begin{equation}
\delta_{\bar{A}A}\ou{\bar{\tau}}{\bar{A}}{\bar{B}i}\delta^{\bar{B}B}=-\ou{\tau}{B}{Ai}.
\end{equation}

To characterize initial data for selfdual gravity, we have to impose the Gauss, vector, and scalar constraints on phase space, which are the Hamiltonian generators of $SL(2,\C)$ frame rotations, spatial diffeomorphisms and hypersurface deformations. In terms of the Ashtekar variables $(\uo{\tilde{E}}{i}{a},\ou{A}{i}{a})$, the constraints are given by
\begin{subequations}
\begin{align}
G_i[\Lambda^i]&=\int_M\Lambda^iD_a\uo{\tilde{E}}{i}{a},\label{contgauss}\\
H_a[N^a]&=\int_MN^a\ou{F}{i}{ab}\uo{\tilde{E}}{i}{b},\label{contvec}\\
H[\utilde{N}]&=\int_M\utilde{N}\uo{\epsilon}{i}{lm}\ou{F}{i}{ab}\uo{\tilde{E}}{l}{a}\uo{\tilde{E}}{m}{b}\label{contscal},
\end{align}
\end{subequations}
where $D_a$ is the $SL(2,\C)$ gauge covariant derivative,  $\ou{F}{i}{ab}=2\partial_{[a}\ou{A}{i}{b]}+\ou{\epsilon}{i}{lm}\ou{A}{l}{a}\ou{A}{m}{b}$ is its curvature, and $\Lambda^i$, $N^a$ and $\utilde{N}$ are smearing functions of compact support ($\utilde{N}$ is a inverse density of weight minus one). Its density weight will play an important role below.

The resulting constraint algebra is first-class. The only non-trivial Poisson commutators among the constraints are given by
\begin{subequations}
\begin{align}
\big\{G_i[\Lambda^i],G_j[\mathrm{M}^j]\big\}&=-8\pi\I G\,G_i\big[[\Lambda,\mathrm{M}]^i\big],\\
\big\{H_a[N^a],H_b[M^b]\big\}&=-8\pi\I G\Big(H_a\big[[N,M]^a\big]-G_i\big[\ou{F}{i}{ab}N^aM^b\big]\Big),\label{VecVecpoiss}\\
\big\{H_a[N^a],H[\utilde{N}]\big\}&=-8\pi\I G \Big(H[\mathcal{L}_{\vec{N}}\utilde{N}]+G_i\big[\uo{\epsilon}{j}{ki}\ou{F}{j}{ab}\uo{\tilde{E}}{k}{a}\utilde{N}N^b\big]\Big),\label{VecHpoiss}\\
\big\{H[\utilde{N}],H[\utilde{M}]\big\}&=+8\pi\I G\,H_a\big[[\utilde{N},\utilde{M}]^a\big].\label{HHpoiss}
\end{align}
\end{subequations}
All other commutators among the constraints vanish identically on the entire kinematical phase space. In here, $[\Lambda,\mathrm{M}]^i=\ou{\epsilon}{i}{jk}\Lambda^j\mathrm{M}^k$ is the Lie bracket with respect to the $\mathfrak{sl}(2,\C)$ Lie algebra and the Lie bracket between vector fields is denoted by $[N,M]^a=N^bD_bM^a-M^bD_bN^a$. Furthermore, $\mathcal{L}_{\vec{N}}\utilde{N}=-(\utilde{N})^2D_a((\utilde{N})^{-1}N^a)$ is the Lie derivative along the vector field $N^a\in TM$ of the density weight $-1$ lapse function $\utilde{N}$. For any two such scalar densities $\utilde{M}$ and $\utilde{N}$ of weight $-1$, the bracket $[\utilde{N},\utilde{M}]^a$ denotes the vector field
\begin{equation}
[\utilde{N},\utilde{M}]^a=\delta^{ij}\uo{\tilde{E}}{i}{a}\uo{\tilde{E}}{j}{b}(\utilde{N}D_b\utilde{M}-\utilde{M}D_b\utilde{N}).
\end{equation}
Notice that $[\utilde{N},\utilde{M}]^a$ is independent of the choice of covariant derivative operator $D_a$, i.e.\ $\utilde{N}D_b\utilde{M}-\utilde{M}D_b\utilde{N}=\utilde{N}\partial_b\utilde{M}-\utilde{M}\partial_b\utilde{N}$ for some fiducial flat derivative $\partial_a:[\partial_a,\partial_b]=0$. Notice also that the constraint algebra remains well-defined even for those field configurations for which the densitized triads are degenerate, i.e.\ $\epsilon^{ijk}\uo{\tilde{E}}{i}{a}\uo{\tilde{E}}{j}{b}\uo{\tilde{E}}{k}{c}=0$.

\section{New lattice regularization of the constraints}\label{sec3}
\subsection{Basic holonomies and fluxes}\label{sec3.1}\noindent
In the following, we consider the discretization of the constraints (\ref{contgauss}, \ref{contvec}, \ref{contscal}) in a small tetrahedron $T\subset M$, whose four corners are located at coordinate values $ X^\mu_I:\sum_{I=1}^4X^\mu_I=0$ with respect to some fiducial coordinate system $\{x^\mu\}$ (see \hyperref[fig1]{figure 1} for an illustration). The origin of this coordinate system is put into the centroid of the tetrahedron. In these coordinates, the tetrahedron is the point set $\{x^\mu\in\R^3:x^\mu=\sum_{I=1}^4p^IX^\mu_I,\sum_{I=1}^4p^I\leq 1,0\leq p^I\}$. In the limit in which the \emph{coordinate lengths} $X^\mu_I$ shrink to zero, i.e.\ $X^\mu_I\rightarrow 0$, the tetrahedron shrinks into its centroid.

To discretize the constraints, we first need to discretise the basic variables on phase space, namely the Ashtekar connection $\ou{A}{i}{a}$ and the electric field $\uo{\tilde{E}}{i}{a}$. To discretize the connection, consider the path $\gamma_I$ from the centroid $c$ of the tetrahedron to the centroid of the $I$-th face. With respect to the fiducial coordinate system, this path is given by the map
\begin{equation}
[0,1]\ni t\mapsto \gamma^\mu_I(t)=-\frac{t}{3} X^\mu_I\in \R^3.\label{pathdef}
\end{equation}
The discretized connection is now simply given by the $SL(2,\C)$ parallel transport with respect to the Ashtekar connection along $\gamma_I$. This is the bulk holonomy
\begin{equation}
h_I=\mathrm{Pexp}\Big(-\int_{\gamma_I} A\Big)\in SL(2,\C).\label{holdef}
\end{equation}
\begin{figure}[h]
\begin{center}
\psfrag{I}{$I$}
\psfrag{J}{$J$}
\psfrag{K}{$K$}
\psfrag{L}{$L$}
\psfrag{f}{\large$f^I$}

\psfrag{h}{$f_{JK}$}
\psfrag{g}{$\gamma_{JI}$}
\psfrag{k}{$\gamma_{J}$}
\hspace{1.5em}\includegraphics[width=0.5\textwidth]{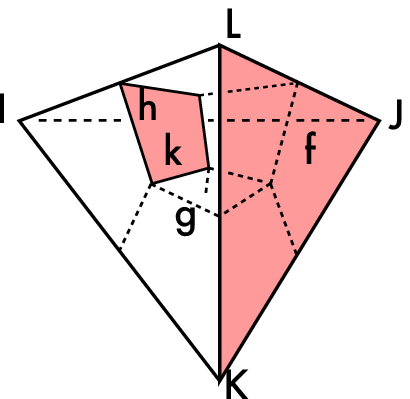}
\end{center}
\caption{Points, paths and surfaces in a tetrahedron: the bulk links $\gamma_I$ connect the centroid of the tetrahedron with the centroid of the $I$-th face $f^I$. The boundary links $\gamma_{IJ}$ lie tangential to the boundary and connect the centroid of $f^I$ with the centroid of $f^J$. The four triangles $f^1,\dots,f^4$ form the boundary of the tetrahedron, the dual faces $f_{JK}$, on the other hand, lie inside the tetrahedron and are bounded by the path $\gamma_K^{-1}\circ\gamma_{JK}\circ\gamma_J$ (this choice implicitly fixes the orientation of $f_{JK}$, while the orientation of $f^I$ is induced from the interior of the tetrahedron). \label{fig1}}
\end{figure}%
To discretize the densitized triad $\uo{\tilde{E}}{i}{a}$, it is useful to assume that the curvature is located \emph{inside} the tetrahedron and the boundary is flat. This implies that there exists a gauge element $g:\partial T\rightarrow SL(2,\C)$ such that the pull-back of the connection to the boundary is
\begin{equation}
\varphi^\ast_{\partial T}\ou{A}{i}{a}\tau_i\equiv\ou{A}{i}{\underleftarrow{a}}\tau_i=g^{-1}\partial_{\underleftarrow{a}}g,\label{bndryflat}
\end{equation}
where  $\varphi^\ast_{\partial T}$ denotes the pull-back to the boundary. If $c_I$ denotes the centroid of the $I$-th face, the parallel transport along the boundary from $c_I$ to $c_J$ is now simply the product
\begin{equation}
\mathrm{Pexp}\Big(-\int_{\gamma_{IJ}} A\Big)=g_J^{-1}g_{I},\label{holdef2}
\end{equation}
where $g_I=g(c_I)$. In \eref{holdef}, the bulk connection $\ou{A}{i}{a}$ is smeared over the links of the tetrahedron. Its conjugate momentum, which is the densitized triad $\uo{\tilde{E}}{i}{a}$, is smeared over the dual faces, which are the four boundary triangles $f^I$. To maintain local gauge invariance, we use the boundary parallel transport and map all free gauge indices into the centre of the tetrahedron. In this way, we obtain the \emph{electric flux},
\begin{equation}
\uo{E}{i}{I}=\int_{f^I}(E_{j})_x\ou{[g^{-1}(x)g_Ih_I]}{j}{i},\label{fluxdef}
\end{equation}
where $E_{j}$ denotes the dual two-form, which is obtained by dualizing the densitized triad with respect to the metric-independent and inverse Levi-Civita tensor density\footnote{Given local coordinates $\{x^\mu\}$, the tensor densities $\utilde{\epsilon}_{abc}$ and $\tilde{\epsilon}^{abc}$ are defined as $\utilde{\epsilon}_{abc}:=\partial_{\mu}\wedge\partial_\nu\wedge\partial_\rho\,\di x^\mu_a\di x^\nu_b\di x^\rho_c$ and $\utilde{\epsilon}^{abc}=\di x^\mu\wedge\di x^\nu\wedge\di x^\rho\partial^a_\mu\partial^b_\nu\partial^c_\rho$.} $\utilde{\epsilon}_{abc}$. Our conventions are
\begin{equation}
E_{jab}=\utilde{\epsilon}_{abc}\uo{\tilde{E}}{j}{c}.
\end{equation}
In addition, $\ou{[h]}{i}{j}$ denotes the adjoint representation of $SL(2,\C)\ni h$,
\begin{equation}
h^{-1}\tau^ih=\ou{[h]}{i}{j}\tau^j,\label{adjnt}
\end{equation}
where $\tau_j$ are the anti-hermitian Pauli matrices \eref{Pauliident}.
\begin{figure}[h]
\begin{center}
\psfrag{I}{\Large$2$}
\psfrag{J}{\Large$1$}
\psfrag{K}{\Large$4$}
\psfrag{L}{\Large$3$}
\psfrag{A}{\Large$u_1{}^a$}
\psfrag{B}{\Large$u_2{}^a$}
\psfrag{C}{\Large$u_3{}^a$}

\hspace{1.5em}\includegraphics[width=0.45\textwidth]{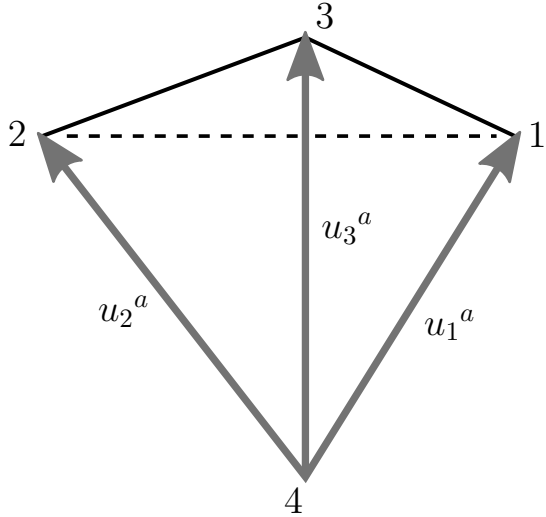}
\end{center}\label{fig2}
\caption{We use a non-orthonormal right-handed basis $(\uo{u}{1}{a},\uo{u}{2}{a},\uo{u}{3}{a})$ of tangent vectors $\uo{u}{1}{a},\dots\in T_{p_4}M$ aligned to three sides of the tetrahedron. The triad is based at one of its corners. In here, this is the fourth corner $p_4$.}%
\end{figure}
\subsection{Regularization of the Gauss constraint}
\noindent The next step is to regularize the constraints. First of all, we consider the Gauss constraint \eref{contgauss}.  Using the non-abelian version of Stokes's theorem, we integrate  Gauss constraint \eref{contgauss} against a smearing function $\Lambda^i$ inside the tetrahedron and obtain (for small curvature) the closure constraint,
\begin{equation}
G_i^T\Lambda^i=\sum_{I=1}^4 \Lambda^iE_i^I\approx\int_{T}\Lambda^i D_a\uo{\tilde{E}}{i}{a}.\label{closd}
\end{equation}

To regularize the other two constraints, it is useful to introduce an adapted triad (a right-handed basis $(\uo{u}{1}{a},\uo{u}{2}{a},\uo{u}{3}{a})$ of $T^\ast M$) at a marked point of the tetrahedron. The simplest choice is to put the origin of the coordinate system at one of the vertices of the tetrahedron (e.g.\ the fourth) and use the three edge vectors pointing to the other three vertices as our basis elements, see \hyperref[fig2]{figure 2}. With respect to our fiducial coordinate system $\{x^\mu\}$ the basis vectors are now given by their coordinate expressions
\begin{equation}
\uo{u}{1}{\mu}=X_1^\mu-X^\mu_4,\quad \uo{u}{2}{\mu}=X_2^\mu-X^\mu_4,\quad\uo{u}{3}{\mu}=X_3^\mu-X^\mu_4.\label{triaddef}
\end{equation}
Given the triadic basis $\uo{u}{\alpha}{a}=\uo{u}{\alpha}{\mu}\partial^a_\mu$, for $\alpha=1,2,3$, we now also have a dual cotriad $\{\ou{u}{\alpha}{a}\}_{\alpha=1,2,3}$ such that
\begin{subalign}
\uo{u}{\alpha}{a}\ou{u}{\alpha}{b}&=\delta^a_b,\label{compltbas}\\
\ou{u}{\alpha}{a}\uo{u}{\beta}{a}&=\delta^\alpha_\beta.
\end{subalign}
In addition, we may now also introduce the scalar densities
\begin{subalign}
(d^3u)&=\tilde{\epsilon}^{abc}\ou{u}{1}{a}\ou{u}{2}{b}\ou{u}{3}{c},\label{densdef}\\
(d^3u)^{-1}&=\utilde{\epsilon}_{abc}\uo{u}{1}{a}\uo{u}{2}{b}\uo{u}{3}{c}.
\end{subalign}
Consider now the flux through the first triangle $f^1$. To leading order in the coordinate size of the tetrahedron,\footnote{Provided the fields do not vary much over the extension of the tetrahedron. Accordingly, the symbol \qq{$\approx$} denotes equality up to subleading terms in a Taylor expansion around the centroid of the tetrahedron, e.g.\ $f(\varepsilon X_I)= f(c)+X^a_I(\partial_a f)(c)+\mathcal{O}(X^2)\approx f(c)$. } we then have
\begin{equation}
\uo{E}{i}{1}\approx -\frac{1}{2}\uo{u}{2}{a}\uo{u}{3}{b}E_{iab}\big|_c=-\frac{1}{2}\utilde{\epsilon}_{abc}\uo{u}{2}{a}\uo{u}{3}{b}\uo{u}{\mu}{c}\ou{u}{\mu}{a}\uo{\tilde{E}}{i}{a}\big|_c=-\frac{1}{2}(d^3 u)^{-1}\ou{u}{1}{a}\uo{\tilde{E}}{i}{a}\big|_c,\label{fluxapprox1}
\end{equation}
where all fields are evaluated at the centroid of the tetrahedron (or at any other point inside $T$). Equation  \eref{fluxapprox} must be true for all three faces $f^1, f^2$ and $f^3$, hence
\begin{equation}
\uo{E}{i}{\alpha}\approx-\frac{1}{2}(d^3 u)^{-1}\ou{u}{\alpha}{a}\uo{\tilde{E}}{i}{a}\big|_c,\quad \alpha=1,2,3\label{fluxapprox}.
\end{equation}
Notice also that the forth missing flux can be expressed in terms of the triad $\uo{E}{i}{\alpha}$. Going back to \eref{closd}, we have
\begin{equation}
\uo{E}{i}{4}=-\sum_{\alpha=1}^3\uo{E}{i}{\alpha}.\label{4flux}
\end{equation}

\subsection{Regularization of the scalar constraint}\noindent
The Gauss, vector and Hamiltonian constraints (\ref{contgauss},\ref{contvec},\ref{contscal}) of selfdual gravity are constructed from tensor densities. To regularize such densities, we proceed as follows. First of all, we note that the integral of the fiducial density $d^3u$ (see \eref{densdef}) over $T$ is nothing but
\begin{equation}
\int_{T}d^3 u=\frac{1}{3!}.\label{d3uInt}
\end{equation}
This is so, because the corners of the tetrahedron are located at the coordinate values $(1,0,0)$, $(0,1,0)$ and $(0,0,1)$ with respect to the three basis vectors $\uo{u}{1}{a}$, $\uo{u}{2}{a}$ and $\uo{u}{3}{a}$. Given the integral of the fiducial density $d^3 u$ along $T$, we obtain the following approximation for the integral of a varying scalar density $\tilde{f}$ of weight one within $T$, namely
\begin{equation}
\int_T \tilde{f} \approx \frac{1}{3!}\big[(d^3 u)^{-1}\tilde{f}\big]_c,\label{Hamapprox1}
\end{equation}
where $c$ is again a point inside $T$ (such as the centroid of the tetrahedron). 

In Ashtekar's self-dual formulation, the scalar constraint  is a density of weight two. The corresponding smearing function $\utilde{N}$ is an inverse density of compact support. To regularize $\utilde{N}$, we take its inverse and integrate the resulting density over the tetrahedron. In other words,
\begin{equation}
N_T:=\Big[\int_T \utilde{N}^{-1}\Big]^{-1}\approx 3! \big[d^3 u\,\utilde{N}\big]_c.\label{Hamapprox2}
\end{equation}

Given \eref{Hamapprox1} and \eref{Hamapprox2}, we obtain \begin{equation}
\int_T\utilde{N}\uo{\epsilon}{i}{lm}\ou{F}{i}{ab}\uo{\tilde{E}}{l}{a}\uo{\tilde{E}}{m}{b}\approx N_T\frac{(d^3 u)^{-2}}{(3!)^2}\uo{\epsilon}{i}{lm}\ou{F}{i}{ab}\uo{\tilde{E}}{l}{a}\uo{\tilde{E}}{m}{b}\Big|_c
\end{equation}
as the leading contribution of the expansion of the smeared scalar constrained in the tetrahedron $T$. The next step is to replace in this expression the densitized triad $\uo{\tilde{E}}{i}{a}$ by the electric fluxes over the four sides of the tetrahedron, see \eref{fluxdef}. We insert the completeness relation \eref{compltbas} and return to our expression for the electric fluxes $E_i^\alpha$, $\alpha=1,2,3$ as given in \eref{fluxapprox}. We obtain
\begin{equation}
\int_T\utilde{N}\uo{\epsilon}{i}{lm}\ou{F}{i}{ab}\uo{\tilde{E}}{l}{a}\uo{\tilde{E}}{m}{b}\approx \frac{1}{9}N_T\uo{\epsilon}{i}{lm}\big(\ou{F}{i}{ab}\uo{u}{\alpha}{a}\uo{u}{\beta}{b}\big)_c\uo{E}{l}{\alpha}\uo{E}{m}{\beta}.\label{Hamapprox3}
\end{equation}
The curvature term has a neat geometric interpretation in terms of \emph{magnetic fluxes} through the dual facets $\{f_{IJ}\}$, see  \hyperref[fig1]{figure 1}. Any such surface $f_{IJ}$ has the geometry of a kite: its four corners are the centroid of the tetrahedron, which is the origin of our fiducial coordinate system $\{x^\mu\}$, the centroids $\uo{c}{I}{\mu}=-\frac{1}{3}X^\mu_I$ and $\uo{c}{J}{\mu}=-\frac{1}{3}X^\mu_J$ of the triangles $f^I$ and $f^J$, in addition to $\uo{c}{KL}{\mu}=-\frac{1}{2}(X^\mu_I+X^\mu_J)$, which is the centroid of the edge\footnote{We are assuming here that $(IJKL)$ is a permutation of $(1234)$.} connecting $X_K^\mu$ and $X_L^\mu$. The integral of a slowly varying two-form $\omega$ over $f_{IJ}$ can be approximated, therefore, by the following  expression,
\begin{equation}
\int_{f_{IJ}}\omega\approx \frac{1}{2}\big(\omega_{ab}\,\uo{c}{I}{a}\uo{c}{KL}{b}+\omega_{ab}\,\uo{c}{KL}{a}\uo{c}{J}{b}\big)_c=\frac{1}{6}\omega_{ab}X^a_IX^b_J\big|_c\equiv\frac{1}{6}\omega_c(X_I,X_J),\label{fIJfluxs}
\end{equation}
where $X^a_I$ are the position vectors $X_I^a=X^\mu_I\partial^a_\mu$ of the four vertices of the tetrahedron.  To regularize the scalar constraint \eref{Hamapprox3}, let us then notice that
\begin{align}\nonumber
\omega_c(X_\alpha,X_\beta)&=\omega_c(X_\alpha-X_4,X_\beta-X_4)+\omega_c(X_4,X_\beta-X_\alpha)=\\
&=\omega_{ab}\uo{u}{\alpha}{a}\uo{u}{\beta}{b}\big|_c+\omega_c(X_4,X_\beta-X_\alpha).\label{fluxrel}
\end{align}
where we reintroduced the triad $\uo{u}{\alpha}{a}$ adapted to the sides of the tetrahedron, see \eref{triaddef}. Let now $F^i[f_{IJ}]$ denote the integral of the curvature two-form over $f_{IJ}$, i.e.
\begin{equation}
F^i[f_{IJ}]=\int_{f_{IJ}}F^i.
\end{equation}
Going back to \eref{fluxrel}, we obtain the important identity
\begin{equation}
\frac{1}{6}\ou{F}{i}{ab}\uo{u}{\alpha}{a}\uo{u}{\beta}{b}\big|_c\approx  F^i[f_{\alpha\beta}]-F^i[f_{4\beta}]-F^i[f_{\alpha 4}].\label{magfluxident}
\end{equation}
We insert this expression back into our regularization for the scalar constraint \eref{Hamapprox3}, which leads us to
\begin{align}\nonumber
\int_T\utilde{N}\uo{\epsilon}{i}{lm}\ou{F}{i}{ab}\uo{\tilde{E}}{l}{a}\uo{\tilde{E}}{m}{b}\approx\frac{2}{3}N_T\Big(&\uo{\epsilon}{i}{lm}F^i[f_{\alpha\beta}]\uo{E}{l}{\alpha}\uo{E}{m}{\beta}+\\
&-\uo{\epsilon}{i}{lm}F^i[f_{4\beta}]\uo{E}{l}{\alpha}\uo{E}{m}{\beta}-\uo{\epsilon}{i}{lm}F^i[f_{\alpha4}]\uo{E}{l}{\alpha}\uo{E}{m}{\beta}\Big),
\end{align}
where the sum is taken over all indices $\alpha,\beta=1,2,3$. The closure constraint \eref{4flux} finally brings this equation into the following neat form
\begin{equation}
\int_T\utilde{N}\uo{\epsilon}{i}{lm}\ou{F}{i}{ab}\uo{\tilde{E}}{l}{a}\uo{\tilde{E}}{m}{b}\approx\frac{2}{3}N_T\uo{\epsilon}{i}{lm}F^i[f_{IJ}]\uo{E}{l}{I}\uo{E}{m}{J},\label{Hamapproxstep}
\end{equation}
where we now sum  over all repeated indices $I,J,K,\dots = 1,\dots,4$ labelling the four sides of the tetrahedron.\medskip

On the lattice, the fundamental configuration variables are the parallel propagators (\ref{holdef}, \ref{holdef2}) and fluxes \eref{fluxdef}. The integral of the curvature two-form along the dual faces $f_{IJ}$ must be approximated, therefore, by parallel propagators. This can be achieved by using the non-abelian Stokes's theorem, which  states that the holonomy around the perimeter of a surface $f$ is the surface ordered exponential of the field strength,
\begin{equation}
\mathrm{Pexp}\Big(-\oint_f A\Big)=\mathrm{Sexp}\Big(-\int_fF\Big)\approx \bbvar{1}-\int_f F^i\tau_i+\dots\label{stokes}
\end{equation}
On the other hand, we now also know from \eref{holdef} and \eref{holdef2} that the holonomy around the perimeter of $f_{IJ}$ is given by the \emph{magnetic fluxes},
\begin{equation}
F_{IJ}:=\mathrm{Pexp}\Big(-\oint_{f_{IJ}} A\Big)=h_J^{-1}g_J^{-1}g_Ih_I\in SL(2,\C).\label{magflux}
\end{equation}
Replacing in \eref{Hamapproxstep} the integral of $F^i$ over $f_{IJ}$ by the holonomy around the boundary, we obtain
\begin{equation}\label{Hdef}
H^TN_T:=\frac{8}{3}N_T\mathrm{Tr}\big(\tau^jF_{IJ}\tau^i\big)\uo{E}{i}{I}\uo{E}{j}{J}\approx\int_T\utilde{N}\uo{\epsilon}{i}{lm}\ou{F}{i}{ab}\uo{\tilde{E}}{l}{a}\uo{\tilde{E}}{m}{b}
\end{equation}
as a regularization of the scalar constraint smeared over the tetrahedron $T$, where we sum over all repeated indices $I,J,=1,\dots,4$ and $i,j=1,2,3$.
\subsection{Regularization of the vector constraint}\noindent
In the previous section, we considered the scalar constraint. Next, we turn to the vector constraint. First of all, we decompose the corresponding vector-valued Lagrange multiplier $N^a$ with respect to the triadic basis \eref{triaddef}, obtaining
\begin{equation}
N^a\big|_c=N^\alpha\uo{u}{\alpha}{a}.
\end{equation}
Using our approximation scheme for scalar densities \eref{Hamapprox1} and the completeness relations \eref{compltbas} for the triadic basis $\{\uo{u}{\alpha}{a}\}_{\alpha=1,2,3}$, we then have
\begin{align}
\int_TN^a\ou{F}{i}{ab}\uo{\tilde{E}}{i}{b}\approx\frac{1}{3!}N^\alpha\ou{F}{i}{ab}\uo{u}{\alpha}{a}\uo{u}{\beta}{b}(d^3 u)^{-1}\ou{u}{\beta}{d}\uo{\tilde{E}}{i}{d}\big|_c.
\end{align}
If we now use the approximation for the electric fluxes \eref{fluxapprox} together with the identity \eref{magfluxident} for the magnetic fluxes, we can rewrite this expression as follows,
\begin{align}\nonumber
\int_TN^a\ou{F}{i}{ab}\uo{\tilde{E}}{i}{b}&\approx-\frac{1}{3}N^\alpha\big(\ou{F}{i}{ab}\uo{u}{\alpha}{a}\uo{u}{\beta}{b}\big)_c\uo{E}{i}{\beta}=\\
&\approx-2 N^\alpha\big(F^i[f_{\alpha\beta}]+F^i[f_{\alpha 4}]-F^i[f_{4\beta}]\big)\uo{E}{i}{\beta},\label{vecapprox1}
\end{align}
where we sum again over all indices $\alpha,\beta=1,2,3$. The final expression \eref{vecapprox1} suggests to introduce a four-vector $N^I\in\R^4$, whose components are given by
\begin{equation}
N^{I=\mu}=N^\mu,\quad N^{I=4}=-\sum_{\mu=1}^3N^\mu.
\end{equation}
In other words, the discretized Lagrange multiplier $N^I\in\R^4$ is not arbitrary, but satisfies a closure constraint as well, namely
\begin{equation}
\sum_{I=1}^4 N^I=0.\label{Nclos}
\end{equation}
If we now also take into account the closure constraint for the electric fluxes, namely \eref{4flux}, we can recast the regularized vector constraint into the following form,
\begin{equation}
\int_TN^a\ou{F}{i}{ab}\uo{\tilde{E}}{i}{b}\approx-2N^I F^i[f_{IJ}]\uo{E}{i}{J}.
\end{equation}
Finally, we have to replace the integral $F^i[f_{IJ}]=\int_{f_{IJ}}F^i$ by the $SL(2,\C)$ group valued magnetic fluxes \eref{magflux}. This leads us to our final expression for the regularized vector constrained in a single tetrahedron, namely,
\begin{equation}
H^T_IN^I:=-4N^I\mathrm{Tr}\big(\tau^jF_{IJ}\big)\uo{E}{j}{J}\approx\int_TN^a\ou{F}{i}{ab}\uo{E}{i}{b},\label{vecdef}
\end{equation}
where we sum over all repeated indices $I,J=1,\dots,4$ for a smearing function $N^I$ that satisfies the closure constraint \eref{Nclos}.
\subsection{Motivation for a second closure constraint}\noindent
In the continuum, the Gauss, vector and scalar constraints (\ref{contgauss}, \ref{contvec}, \ref{contscal}) satisfy a closed algebra. The kinematical phase space of field configurations $(\uo{\tilde{E}}{i}{a},\ou{A}{i}{a})$ has $3\times 3\times 2$ complex dimensions (per point), and there are $3+3+1=7$ complex constraints, which are all first class. The resulting reduced phase space has $18-2\times 7=4$ complex dimensions, which describe the two complex degrees of freedom of self-dual gravity (the two polarisations of the self-dual graviton).

In the previous section, we introduced a lattice regularization for the Gauss \eref{closd}, scalar \eref{Hdef} and vector constraint \eref{vecdef} in a single tetrahedron. Do we have a chance to recover the right degrees of freedom at the discretized level? Suppose, for example that the discrete constraints (\ref{closd}, \ref{Hdef}, \ref{vecdef}) are all first-class, which would give us seven (complex) constraints for each tetrahedron. How many dimensions would the resulting reduced phase space have? In the continuum, the phase space is equipped with the symplectic potential
\begin{equation}
\Theta_M(\delta)=8\pi \I G\int_M\uo{\tilde{E}}{i}{a}\delta\ou{A}{i}{a}.\label{thetacont}
\end{equation}
At the discretized level, the contribution to the symplectic potential from each tetrahedron is\footnote{As fas as the Hamiltonian analysis for a \emph{single} tetrahedron is concerned, the four edge modes (boundary holonomies) $g_I\in SL(2,\C)$ are mere spectators, and Poisson commute with all bulk variables $(\uo{E}{i}{I},h_I)\in\mathfrak{sl}(2,\C)\times SL(2,\C)$. Their own phase space and canonical momenta become relevant only once we consider the glueing of adjacent tetrahedra, see \hyperref[sec4]{section 4}.}
\begin{equation}
\Theta_T(\delta)=16\pi\I G\sum_{I=1}^{4}\uo{E}{i}{I}\mathrm{Tr}\big(\tau^ih^{-1}_I\delta h_I\big),\label{thetadisc}
\end{equation}
see e.g.\ \cite{twist3}. The fundamental configuration variables of the discretized theory are the bulk holonomies $h_I$, which are parametrized by $4\times 3=12$ complex numbers, the resulting lattice phase space (see section \hyperref[sec3.1]{section 3.1}) has 24 complex dimensions. The naive counting would then give $24-2\times 7=10$ physical degrees of freedom per tetrahedron, which does not match the theory in the continuum. There seem to be three unphysical lattice degrees of freedom. That there are such spurious lattice modes has a simple explanation. The symplectic potential for the lattice phase space \eref{thetadisc} treats the variables assigned to all four directions $I=1,\dots,4$ in the tetrahedron as independent. This contradicts the theory in the continuum: for small curvature, we can approximate the parallel propagators as
\begin{equation}
h_I=\bbvar{1}+\frac{1}{3}\ou{A}{i}{a}X^a_I\big|_c+\dots,
\end{equation}
see \eref{holdef} and \eref{holdef}. In our derivations of the scalar and vector constraint \eref{vecdef} and \eref{Hdef}, we implicitly assumed that 
\begin{equation}
\sum_{I=1}^4 X^a_I=0.\label{Xclos}
\end{equation}
Therefore, already at the kinematical level, the bulk holonomies $h_I$ and fluxes $\uo{E}{i}{I}$ should not be seen as completely independent variables. Going from \eref{thetacont} to \eref{thetadisc} there are $2\times 3$ additional spurious modes appearing. An additional closure constraint seems to be missing that should impose an appropriate version of \eref{Xclos}.

We now need a candidate for the missing constraint. Consider the following \emph{dressed closure constraint}
\begin{equation}
\sum_{K=1}^4G_k^{(K)}:=\frac{1}{4}\sum_{I,K=1}^4\ou{[F_{KI}]}{i}{k}\uo{E}{i}{I},\label{clos2}
\end{equation}
where $\ou{[h]}{i}{j}$ denotes again the adjoint representation \eref{adjnt} of $SL(2,\C)$ and $F_{IJ}$ is the magnetic flux \eref{magflux}. For small curvature, the $SL(2,\C)$-valued magnetic flux $F_{IJ}$ can be replaced to good approximation by the holonomy
\begin{equation}
F_{IJ}=\mathrm{Pexp}\Big(-\oint_{f_{IJ}}A\Big)=\bbvar{1}-\int_{f_{IJ}}F^i\tau_i+\dots.
\end{equation}
For slowly varying curvature, the integral of the curvature two-form over the dual faces $f_{IJ}$ can be approximated by the components of the curvature two-form $\ou{F}{i}{ab}$ contracted with the position vectors $X_I^a$ and $X_J^b$
\begin{equation}
\int_{f_{IJ}}F^i\approx\frac{1}{6} \ou{F}{i}{ab}X_I^aX_J^b\big|_c,
\end{equation}
see \eref{fIJfluxs}. This in turn allows us to expand equation \eref{clos2} for small curvature. We obtain
\begin{equation}
\sum_{K=1}^4G_k^{(K)}= \sum_{I=1}^4\uo{E}{i}{I}-\frac{1}{24} \sum_{I,K=1}^4\ou{\epsilon}{i}{jk}\ou{F}{j}{ab}X_I^aX_K^b\big|_c\uo{E}{i}{I}+\dots
\end{equation}
The leading order of this expression returns the usual closure constraint \eref{closd}. For generic configurations of  $\ou{F}{i}{ab}$ the next to leading order vanishes if and only if \eref{Xclos} is satisfied. This observation is a first indication to add the \emph{dressed closure constraint} \eref{clos2} to our set of constraints \eref{closd}, \eref{Hdef}, \eref{vecdef}. In the next section, we will find another such indication. The resulting constraint algebra closes under the Poisson bracket.\footnote{There are non-trivial \emph{structure functions}, such that the resulting constraint algebra is not a Lie algebra.} At the kinematical level, there are $4\times 3\times 2=24$ complex phase space dimensions. In addition, there are $3+3+1+3=10$ constraints removing $2\times 10=20$ complex directions in phase space. The resulting reduced phase space describes, therefore, $2$ complex degrees of freedoms that we identify with the two complex degrees of freedom of the selfdual graviton at the discretized level. Notice also that in those special \emph{Regge-like configurations} in which the holonomy $F_{IJ}=\mathrm{Pexp}(-\oint_{f_{IJ}}A)$ around each facet $f_{IJ}$ preserves the flux $E^I$, i.e.\ $\ou{[F_{IJ}]}{j}{i}\uo{E}{j}{J}=\uo{E}{i}{J}$, the dressed closure constraint \eref{clos2} reduces to the usual discretized Gauss constraint \eref{closd}. Thus, for Regge-like configurations, no additional constraint appears.

\section{Phase space for a single cell, closed constraint algebra}\label{sec4}
\subsection{Holonomy flux algebra\label{sec4.1}}\noindent
The Poisson commutation relations for the Ashtekar variables \eref{contpoiss} determine the commutation relations for the corresponding smeared variables, namely the bulk holonomies \eref{holdef} and electric fluxes \eref{fluxdef}. A straight-forward calculation gives,\footnote{A subtlety arises from the fact that the bulk holonomies terminate at the intersection with the triangles, which requires to provide a definition for what is meant by the singular integral $\int_{0}^1\di u\delta(u)$, see \cite{thiemann} for further details. }
\begin{subequations}
\begin{align}
\big\{\ou{[h_I]}{A}{C},\ou{[h_J]}{B}{D}\big\}&=0,\label{hh}\\
\big\{\uo{E}{i}{I},\ou{[h_J]}{A}{B}\big\}&=-8\pi\I G\delta^I_J \ou{[h_I]}{A}{C}\ou{\tau}{C}{Bi},\label{EH}\\
\big\{\uo{E}{i}{I},\uo{E}{j}{J}\big\}&=-8\pi\I G \delta^{IJ}\uo{\epsilon}{ij}{k}\uo{E}{k}{I},\label{EE}
\end{align}
\end{subequations}
which may be derived also directly from \eref{thetadisc}, see \cite{twist3,komplexspinors}. On the resulting $2\times3\times4=24$ complex-dimensional phase space, we need to impose the regularized versions of the Gauss, vector and scalar constraint, see \eref{closd}, \eref{Hdef}, \eref{vecdef}, in addition to the dressed closure constraint \eref{clos2}. To this goal, we now need to calculate the algebra of constraints.

Before we continue, a few further comments. When introducing the holonomies, see \eref{holdef} and \eref{holdef2}, we made an implicit restriction on the allowed boundary data: the pull-back of the connection to the boundary is flat. This restriction implies a relation among the magnetic fluxes, namely
\begin{equation}
\forall I,J,K:F_{KJ}F_{IK}=F_{IJ},\label{Fprod}
\end{equation}
which means that only three out of the six group elements $F_{IJ}\in SL(2,\C)$ are functionally independent.\footnote{This makes a lot of sense geometrically, because in three dimensions there can only be three linearly independent bivectors (planes), and the holonomies $F_{IJ}$ measure the magnetic flux through the planes $f_{IJ}$, see \hyperref[fig1]{figure 1}.} If we take this condition into account, we see that the dressed closure constraint simplifies \eref{clos2}. One of the sums can be dropped, and the constraint simplifies to
\begin{equation}
G_k^{(K)}:=\sum_{I=1}^4\ou{[F_{KI}]}{i}{k}\uo{E}{i}{I}=\ou{[F_{K1}]}{i}{k}G_i^{(1)}\stackrel{!}{=}0,
\end{equation}
for some arbitrary $K\in\{1,\dots,4\}$.

Notice also that the bulk holonomies $h_I, I=1,\dots,4$ enter the definition of the constraints only through the magnetic fluxes $F_{IJ}\in SL(2,\C)$. It is therefore more convenient to use the commutation relations between electric and magnetic fluxes rather than \eref{EH}, which are given by
\begin{equation}
\big\{F_{IJ},\uo{E}{l}{L}\big\}=8\pi\I G\big(\delta^L_IF_{IJ}\tau_l-\delta^L_J\tau_lF_{IJ}\big).\label{EF}
\end{equation}

 Finally, let us briefly summarize what we have done so far. In a tetrahedron, which is small against the curvature scale at which the fields fluctuate, the Gauss \eref{contgauss}, vector \eref{contvec} and scalar \eref{contscal} constraints of selfdual gravity can be written to good approximation in terms of electric and magnetic fluxes alone, see \eref{closd}, \eref{Hdef} and \eref{vecdef}. In addition, there is one more constraint, that does not appear in the continuum, namely the dressed closure constraint \eref{clos2}. The role of this additional constraint is to remove unphysical degrees of freedom that appeared by introducing the discretization. That there are such unphysical lattice degrees of freedom follows from a simple counting argument: the kinematical phase space in the continuum \eref{thetacont} has $2\times3\times 3=18$ complex dimensions per point, the lattice phase space for a tetrahedron has $2\times 3\times 4=24$ complex dimensions, hence we should add three additional complex (first class) constraints to remove the spurious directions from phase space. As a candidate for the  missing constraint, we consider the dressed closure constraint. In summary, the list of constraints is given by
\begin{subalign}
&\text{closure constraint:\;}  &&\hspace{-0.7em}G_i[\Lambda^i]=\sum_{I=1}^4\Lambda^i\uo{E}{i}{I}\stackrel{!}{=}0,\;\forall\Lambda^i\in\R^3,\label{Gcons}\\
&\text{dressed closure:\;}  &&\hspace{-0.7em}G^{(K)}_i[\mathrm{M}^i]=\sum_{I=1}^4\mathrm{M}^k\ou{[F_{KI}]}{i}{k}\uo{E}{i}{I}\stackrel{!}{=}0,\;\forall\Lambda^i\in\R^3,\label{DGcons}\\
&\text{vector constraint:\;} 
 &&\hspace{-0.7em}H_I[N^I]=-4\sum_{I,J=1}^4N^I\mathrm{Tr}\big(F_{IJ}\tau^j\big)\uo{E}{j}{J}\stackrel{!}{=}0,\;\forall N^I:\sum_{I=1}^4N^I=0,\label{HIcons}\\
&\text{scalar constraint:\;} &&\hspace{-0.7em} H=\frac{8}{3}\sum_{I,J=1}^4\mathrm{Tr}\big(\tau^iF_{JI}\tau^j\big)\uo{E}{i}{I}\uo{E}{j}{J}\stackrel{!}{=}0.\label{Hcons}
\end{subalign}
The next task ahead is to demonstrate that the corresponding constraint algebra closes under the Poisson brackets \eref{hh}, \eref{EE} and \eref{EF}.
\subsection{Closure constraint}\noindent
Given the commutation relations for the electric fluxes \eref{EE}, it immediately follows that the closure constraint \eref{closd} is the regenerator of  $SL(2,\C)$ gauge transformations at the centre of the tetrahedron. The relevant Poisson brackets are
\begin{subalign}
\big\{G_i,G_j\big\}&=-8\pi\I G\uo{\epsilon}{ij}{k}G_k,\\
\big\{G_i,G_j^{(K)}\big\}&=-8\pi\I G\uo{\epsilon}{ij}{k}G_k^{(K)},\\
\big\{G_i, H_I[N^I]\big\}&=0,\\
\big\{G_i, H\big\}&=0.
\end{subalign}
Thus, the closure constraint \eref{Gcons} weakly commutes will all other constraints. 
\subsection{Dressed closure constraint}\noindent
Next, we consider the dressed closure constraint \eref{DGcons}. To simplify the calculation, it is useful to consider first its Hamiltonian vector field as it acts on the electric and magnetic fluxes. The Poisson brackets between the constraint and the electric fluxes is given by
\begin{equation}
\big\{G_l^{(K)},\uo{E}{i}{I}\big\}=-8\pi\I G\delta^I_K\uo{\epsilon}{li}{n}G_n^{(K)},\label{DGE}
\end{equation}
which follows immediately from \eref{EE} and \eref{EF}. The magnetic fluxes, on the other hand, Poisson commute with the dressed closure	constraint,
\begin{align}
\nonumber\big\{G_i^{(K)},F_{IJ}\big\}&=\sum_{L=1}^4\Lambda^i\ou{[F_{KL}]}{l}{i}\big\{\uo{E}{l}{L},F_{IJ}\big\}=\\
\nonumber&=-8\pi\I G\Lambda^i\Big[\ou{[F_{KI}]}{l}{i}F_{IJ}\tau_l-\ou{[F_{KJ}]}{l}{i}\tau_lF_{IJ}\Big]=\\
&=-8\pi\I G\lambda^i\big[F_{IJ}F_{KI}\tau_iF_{IK}-F_{KJ}\tau_iF_{JK}F_{IJ}\big]=0,\label{DGF}
\end{align}
where we used in the last step the product identity \eref{Fprod} for the magnetic fluxes and $\ou{[h]}{i}{j}$ denotes again the adjoint representation of $SL(2,\C)$, see \eref{adjnt}.

The equations \eref{DGE} and \eref{DGF} tell us that the electric and magnetic fluxes  weakly Poisson commute\footnote{I.e.\ the Poisson brackets vanish modulo other constraints.} with the dressed closure constraint \eref{DGcons}. All our constraints can be expressed in terms of polynomials of the electric and magnetic fluxes alone. Therefore, the dressed closure 
constraint weakly commutes with all other constraints. In particular,
\begin{subalign}
\big\{G_i^{(K)}&, H_I[N^I]\big\}=32\pi\I G \sum_{I=1}^4N^I\mathrm{Tr}\big(F_{IK}\tau^l\big)\uo{\epsilon}{il}{n}G_N^{(K)},\\
\big\{G_i^{(K)}&, H\big\}=-\frac{2^7\pi\I G}{3}\sum_{L=1}^4\mathrm{Tr}\big(\tau^lF_{KL}\tau^m\big)\uo{E}{l}{L}\uo{\epsilon}{im}{n}G_n^{(K)}.
\end{subalign}
To summarise, the dressed closure constraint \eref{DGcons} commutes with all other constraints as well.
\subsection{Vector\,--\,vector bracket}\noindent
The basic commutation relations for the electric and magnetic fluxes \eref{EE} and \eref{EF} imply the Poisson bracket between two vector constraints,
\begin{align}\nonumber
\big\{H_I[N^I],H_J[M^J]\big\}=&2^7\pi\I GN^IM^J\Big[
\mathrm{Tr}\big(F_{IL}\tau^i\big)\mathrm{Tr}\big(F_{JL}\tau_i\tau^l\big)-\mathrm{Tr}\big(F_{IJ}\tau^i\big)\mathrm{Tr}\big(F_{JL}\tau_i\tau^l\big)+\\
&
+\mathrm{Tr}\big(F_{JI}\tau^i\big)\mathrm{Tr}\big(F_{IL}\tau_i\tau^l\big)
-\mathrm{Tr}\big(F_{JL}\tau^i\big)\mathrm{Tr}\big(F_{IL}\tau^l\tau_i\big)]\uo{E}{l}{L}\label{vecvec1}
\end{align}
To simplify our notation, we have adopted a summation convention, where we sum over all repeated indices $I,J,K,\dots=1,\dots,4$ and $i,j,k,\dots=1,2,3$. For any complex matrix $X\in\C^2\otimes (\C^2)^\ast$ and $h\in SL(2,\C)$, we now have the identity
\begin{equation}
\mathrm{Tr}(h\tau_i)\mathrm{Tr}(X\tau^i)=\frac{1}{4}\Big(\mathrm{Tr}(h^{-1}X)-\mathrm{Tr}(hX)\Big).\label{pauli2}
\end{equation}
This idenity brings the expression for the Poisson bracket \eref{vecvec1} into the following form,
\begin{align}\nonumber
\big\{H_I[N^I],H_J[M^J]\big\}=&2^5\pi\I GN^IM^J\Big[\mathrm{Tr}\big(F_{JL}F_{LI}\tau^l\big)-\mathrm{Tr}\big(F_{JL}F_{JI}\tau^l\big)+\mathrm{Tr}\big(F_{JL}F_{IJ}\tau^l\big)+\\
&+\mathrm{Tr}\big(F_{IL}F_{IJ}\tau^l\big)-\mathrm{Tr}\big(F_{IL}F_{JI}\tau^l\big)-\mathrm{Tr}\big(F_{JL}F_{II}\tau^l\big)\Big]\uo{E}{l}{L}.
\end{align}
We now need to show that the right hand side is again a sum of constraints. To this goal, consider first the following algebraic identity,
\begin{equation}
\sum_{J=1}^4F_{JI}\tau^i\uo{E}{i}{J}=\frac{1}{4}H_I\bbvar{1}+\sum_{J=1}^4\tau^iF_{IJ}\uo{E}{i}{J},\label{vecident}
\end{equation}
which follows from the Pauli identity \eref{Pauliident} and the definition of $H_I$, namely\footnote{Notice that the vector constraint \eref{HIcons} implies $\sum_IH_IN^I=0$ only for those vectors $N^I$ for which $\sum_{I+1}^4N^I=0$, hence $H_I\neq 0$ for generic configurations on phase space.}
\begin{equation}
H_I=-4\sum_{J=1}^4\mathrm{Tr}\big(F_{IJ}\tau^i\big)\uo{E}{i}{J}.
\end{equation}
If we now also use the product identity \eref{Fprod} for the magnetic fluxes and take into account that the smearing functions $N^I$ and $M^I$ satisfy $\sum_IN^I=\sum_IM^I=0$, we obtain 
\begin{align}\nonumber
\big\{H_I[N^I],H_J[M^J]\big\}=&2^5\pi\I G\sum_{I,J=1}^4N^IM^J\Big[\mathrm{Tr}\big(F_{IJ}\tau^m\big)\sum_{L=1}^4\ou{[F_{JL}]}{l}{m}\uo{E}{l}{L}+\\&-\mathrm{Tr}\big(F_{IJ}\tau^l\big)\sum_{L=1}^4\uo{E}{l}{L}
+\frac{1}{4}\mathrm{Tr}\big(F_{IJ}\big)H_J-\frac{1}{4}\mathrm{Tr}\big(F_{IJ}\big)H_I\Big].\label{vecvec3}
\end{align}
The first and second term on the right hand side return the dressed and undressed closure constraints, \eref{Gcons} and \eref{DGcons}. The last two terms are again proportional to the vector constraint. This can be seen as follows. Define the following vector,
\begin{equation}
[N,M]^I=\sum_{J=1}^4\Big(N^J\mathrm{Tr}(F_{JI})M^I-M^J\mathrm{Tr}(F_{JI})N^I\Big).\label{vecvecbrack}
\end{equation}
Since $\mathrm{Tr}(F_{IJ})=\mathrm{Tr}(F_{JI})$, we also have $\sum_{I=1}^4[N,M]^I=0$, which implies that the bracket $[N,M]^I$ defines again a discrete smearing function for the vector constraint, since it satisfies the closure condition that any such smearing function ought to obey, see \eref{HIcons}. Going back to the definition of the constraints, we thus see that the right hand side of \eref{vecvec3} is again a sum of constraints. More explicitly,
\begin{align}\nonumber
\big\{H_I[N^I],H_J[M^J]\big\}=-2^5\pi\I G\bigg[\sum_{I,J=1}^4N^IM^J\mathrm{Tr}(F_{IJ}\tau^l)&\big(G_l^{(J)}+G_l\big)
+\\
&+\frac{1}{4}\sum_{I=1}^4H_I[N,M]^I\bigg].\label{vecvec4}
\end{align}
\subsection{Vector\,--\,scalar bracket}\noindent
Finally, let us complete the calculation of the constraint algebra and turn to the Poisson brackets between the scalar and vector constraints. The elementary Poisson brackets for the electric and magnetic fluxes are given in \eref{EE} and \eref{EF}. A straightforward calculation yields
\begin{align}\nonumber
\big\{H_I[N^I],H\big\}=-\frac{2^9\pi\I G}{3}N^I\Big[&\mathrm{Tr}\big(F_{IJ}\tau^i\big)\mathrm{Tr}\big(F_{MJ}\tau^m\tau_i\tau^l\big)\nonumber\\
\nonumber&+\mathrm{Tr}\big(F_{JI}\tau^l\tau_i\big)\mathrm{Tr}\big(F_{MI}\tau^m\tau^i\big)+\\
&+\mathrm{Tr}\big(F_{IJ}\tau^l\tau_i\big)\mathrm{Tr}\big(F_{MJ}\tau^m\tau^i\big)\Big]\uo{E}{l}{J}\uo{E}{m}{M},
\end{align}
where we sum again over all repeated indices $I,J,\dots=1,\dots,4$ and $i,j,k=1,\dots,3$. Consider then the following identity, which is satisfied for any $h\in SL(2,\C)$ and any complex $2\times 2$ matrices $\ou{X}{A}{B}\in \C^2 \otimes(\C^2)^\ast$, namely
\begin{equation}
\mathrm{Tr}\big(h\tau_l\tau_i\big)\mathrm{Tr}\big(X\tau^i\big)=-\frac{1}{4}\Big(\mathrm{Tr}\big(h^{-1}X\tau_l\big)+\mathrm{Tr}\big(Xh\tau_l\big)\Big),\label{pauli3}
\end{equation}
which is a consequence of the fundamental Pauli identity \eref{Pauliident}. With both \eref{pauli2} and \label{pauli3}, we now get
\begin{align}\nonumber
\big\{H_I[N^I],H\big\}=-\frac{2^7\pi\I G}{3}N^I\Big[&\mathrm{Tr}\big(F_{JI}\tau^lF_{MJ}\tau^m\big)-\mathrm{Tr}\big(\tau^lF_{IJ}F_{MI}\tau^m\big)\big)+\\
&-\mathrm{Tr}\big(F_{JI}\tau^lF_{MI}\tau^m\big)+\mathrm{Tr}\big(
\tau^lF_{JI}F_{MJ}\tau^m\big)\Big]\uo{E}{l}{J}\uo{E}{M}{M}.
\end{align}
Next, we use the product identity for the magnetic fluxes, \eref{Fprod}, in addition to the algebraic identity \eref{vecident} to simplify this expression further. If we also take into account that the vector $N^I$ satisfies the closure constraint \ $\sum_{I=1}^4N^I=0$, we arrive at the following expression
\begin{align}\nonumber
\big\{H_I[N^I],H\big\}=-\frac{2^7\pi\I G}{3}&N^I\Big[\mathrm{Tr}\big(F_{MI}\tau^n\tau^m\big)\ou{[F_{MJ}]}{l}{n}\uo{E}{l}{J}\uo{E}{m}{M}+\\
&-\frac{1}{4}H_I\mathrm{Tr}\big(F_{MI}\tau^m\big)\uo{E}{m}{M}+\mathrm{Tr}\big(\tau^lF_{MI}\tau^m\big)\uo{E}{l}{J}\uo{E}{m}{M}\Big].\label{HIHbrack}
\end{align}
where $\ou{[h]}{i}{j}$ denotes again the adjoint representation \eref{adjnt} and we sum over all repeated indices.
 We now need to convince ourselves that the right hand side vanishes provided our set of constraints (\ref{Gcons}, \ref{DGcons}, \ref{HIcons}, \ref{Hcons}) is satisfied. First of all we note that the first term returns the dressed closure consteraint \eref{Gcons} and the last term is proportional to the ordinary closure constraint \eref{Gcons}. The second term, on the other hand, is again proportional to the vector constraint. To see that this is indeed the case, we have to first convince ourselves that the corresponding multiplier satisfies the closure constraint \eref{Nclos}. Equation \eref{HIHbrack} suggests to introduce the following field-dependent multiplier $H\triangleright N^I$, which is defined for any $N^I:\sum_{I=1}^4 N^I=0$ by 
 \begin{equation}
 H\triangleright N^I\equiv\begin{pmatrix}H\triangleright N^1\\H\triangleright N^2\\H\triangleright N^3\\H\triangleright N^4 \end{pmatrix}:=\begin{pmatrix}H_1 N^1\\H_2 N^2\\H_3 N^3\\H_4 N^4 \end{pmatrix}.\label{HNI}
 \end{equation}
 A short moment of reflection reveals that $H\triangleright N^I$ satisfies  again  the closure constraint \eref{Nclos} for the smearing functions $N^I$ provided the vector constraint is satisfied, i.e.
\begin{equation}
\sum_{I=1}^4H\triangleright N^I=\sum_{I=1}^4 H_IN^I=H_I[N^I]\stackrel{c}{=}0,
\end{equation}
where the symbol \qq{$\stackrel{c}{=}$} denotes equality up to terms that vanish provided the constraints (\ref{Gcons}, \ref{DGcons}, \ref{HIcons}, \ref{Hcons}) are satisfied.

Going back to the definition of the discretized vector \eref{HIcons}, closure \eref{Gcons} and dressed closure constraint \eref{DGcons}, we can now give our final expression for the Poisson bracket between the discretized vector and scalar constraints,
\begin{align}\nonumber
\big\{H_I[N^I],H\big\}=-\frac{2^7\pi\I G}{3}&\Big[\sum_{I,J=1}^4N^I\mathrm{Tr}\big(\tau^m\tau^lF_{IJ}\big)\uo{E}{m}{J}G_l^{(J)}+\\
&+\sum_{I,J=1}^4N^I\mathrm{Tr}\big(\tau^l\tau^mF_{IJ}\big)\uo{E}{m}{J}G_l-\frac{1}{16}H_I[H\triangleright N^I]\Big].\label{HIHbrack2}
\end{align}
The right hand side is a sum over constraints, hence the Poisson bracket $\{H_I[N^I],H\}$ weakly vanishes.

Let us briefly summarize the section before discussing the main open task ahead. In this section, we considered the discretized Gauss, vector and scalar constraint in addition to the dressed closure constraint,  and proved that all the constraints (there are $3+3+1+3=10$ of them) commute among themselves. Since there are structure functions rather than structure constants, the resulting Poiusson commutation relations do not define a Lie algebra. This becomes particularly obvious when considering the \emph{field dependent smearing functions} \eref{vecvecbrack} and \eref{HNI} that enter the right hand side of the Poisson brackets between the vector and scalar constraints, see \eref{vecvec4} and \eref{HIHbrack2}. Our result neatly mirrors the situation in the continuum, where the Poisson brackets among the scalar and vector constraint closes, but does not define a Lie algebra, since there are non-trivial structure functions, see e.g.\ \eref{VecHpoiss}.\medskip

The main open problem ahead is to generalize the results beyond a single simplex, to introduce, in other words, a prescription for how to glue neighbouring tetrahedra in such a way that we generate a triangulation of the spatial manifold, see \hyperref[fig3]{figure 3}. Solving this problem amounts to constructing a Hamiltonian description for self-dual gravity on a simplicial lattice. Although the problem may be difficult and tedious, the basic strategy is clear. The starting point is the discretized action on a single tetrahedron, which is the sum of the symplectic potential \eref{thetadisc} and the constraints \eref{Gcons}, \eref{DGcons}, \eref{HIcons}, \eref{Hcons},
\begin{equation}
 S[\underline{E},\underline{h},\underline{N},\underline{g}]=\int_{\mathbb{R}}\di t\left(\Theta_{\underline{E},\underline{h}}\left(\frac{\di}{\di t}\right)-C_A(\underline{E},\underline{h},\underline{g})N^A\right),
\end{equation}
where $\Theta(\delta)=16\pi\I\, G\,\mathrm{Tr}(E^I h_{I}\delta h_I^{-1})$ is the symplectic potential and  $N^A$ is the collection of Lagrange multipliers for the constraints $C_A\equiv(G_i,G_i^{(K)},H,H_I)$. Notice that the constraints depend as functions not only on the discretized phase space variables, namely fluxes $\underline{E}=(E^1,\dots)\in\mathfrak{sl}(2,\mathbb{C})^4$ and holonomies $\underline{h}=(h_1,\dots)\in SL(2,\mathbb{C})^4$, but they also depend on additional group variables $\underline{g}=(g_1,\dots)\in SL(2,\mathbb{C})^4$. These auxiliary variables, which are reminiscent of gravitational edge modes, parametrize the flat boundary connection \eref{bndryflat}. The introduction of these boundary fields was necessary to replace the curvature tensor in a single tetrahedron by the magnetic fluxes $F_{IJ}\in SL(2,\mathbb{C})$, see \eref{magflux}. To obtain the equations of motion for an isolated tetrahedron, the boundary condition $\delta g_I=0$ is fixed. When we glue adjacent regions, the boundary conditions are relaxed and replaced by a gluing condition. For each edge $e$ that connects two adjacent tetrahedra, we impose the constraint ${g_{I_{s(e)}}}=g_{I_{t(e)}}$  between the \emph{source} and \emph{target} nodes of the underlying edge $e=\gamma_{I_{t(e)}}^{-1}\circ\gamma_{I_{s(e)}}$, which is dual to the face $f^{I_{s(e)}}=[f^{I_{t(e)}}]^{-1}$. Assuming a triangulation $\Delta$ without a boundary, the resulting constrained action is given by
\begin{equation}
 S_\Delta[\underline{E},\underline{h},\underline{N},\underline{\lambda}] = \sum_{T\in\Delta_3} S[\underline{E}_T,\underline{h}_T,\underline{N}_T,\underline{g}_T]-\sum_{e\in\Delta_1^\ast}\lambda_e^i
\mathrm{Tr}(\tau_ig_{s(e)}g_{t(e)}^{-1}),\label{cpldactn}
\end{equation}
where the sum is taken over tetrahedra $T$ and edges $e$ connecting adjacent nodes. Each node is dual to a tetrahedron and the second sum goes over next neighbours.
Although the action looks fairly innocent, we expect it to define a highly non-trivial mechanical system with second-class constraints. This expectation is justified, because there are no canonical momenta conjugate to the boundary variables $\underline{g}$. Hence there are additional constraints besides the Gauss, closure, vector and Hamiltonian constraints $C_A=0$ at each lattice site, which, taken by themselves, would weakly Poisson commute among themselves. The additional constraints are the gluing conditions $g_{I_{s(e)}}=g_{I_t(e)}$ for each edge and   further secondary constraints. Common experience with similar mechanical systems  suggest that there will be both first-class and second-class constraints. The additional constraints arise from the stationary points of the action with respect to the variation of the now internal boundary fields $\underline{g}$. The presence of second-class constraints would alter the canonical commutation relations replacing the Poisson bracket by the Dirac bracket. It would be a surprise if this alteration would not affect the commutation relations between the Hamiltonian and vector constraints at \emph{neighbouring} lattice sites, rendering the algebra local rather than ultra-local. This would mirror the situation in the continuum, where the Poisson bracket between localized scalar and vector constraints involves first derivatives of the Dirac delta distribution rather than the bare delta distribution itself, see e.g.\ \eref{VecHpoiss}, \eref{HHpoiss}.
\begin{figure}[h]
\begin{center}
\psfrag{I}{$I$}
\psfrag{J}{$J$}
\psfrag{K}{$K$}
\psfrag{L}{$L$}
\psfrag{T}{\Large$T$}
\psfrag{f}{\large$f^I$}
\psfrag{g}{\large$\gamma_{IJ}$}
\psfrag{In}{$I'$}
\psfrag{Jn}{$J'$}
\psfrag{Kn}{$K'$}
\psfrag{Ln}{$L'$}
\psfrag{Tn}{\Large$T'$}
\psfrag{fn}{\large$f^{I'}$}
\psfrag{gn}{$\gamma_{I'L'}$}
\psfrag{h}{\large$h_{TT'}$}
\hspace{1.5em}\includegraphics[width=0.9\textwidth]{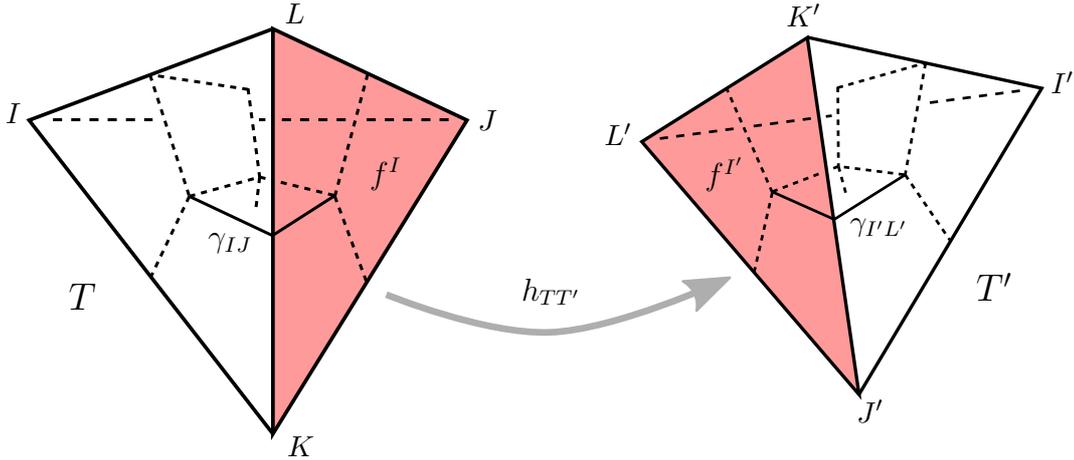}
\end{center}
\caption{Two adjacent tetrahedra $T$ and $T'$ are glued along the triangle $f^I=(f^{I'})^{-1}$ between. The connection at the interface is flat. The transition function from $T$ to $T'$ is given by the $SL(2,\C)$ gauge element $h_{TT'}=g_{I'}^{-1}g_I$.\label{fig3}}
\end{figure}%
\section{Outlook and discussion\label{sec5}}
\noindent There are two main results in this article. The \emph{first result} was developed in \hyperref[sec3]{section 3}, where we found a new lattice regularization of the constraints for self-dual gravity. The construction was motivated by the relatively simple form that the constraints assume in terms of Ashtekar's connection variables. The resulting expressions for the discretized Gauss, vector and Hamilton constraint were given in \eref{closd}, \eref{Hdef} and \eref{vecdef}. At the discretized level, the kinematical phase space for each tetrahedron is $T^\ast SL(2,\mathbb{C})^4$, which has $2\times 3\times 4=24$ complex dimensions. This observation creates a tension with the situation in the continuum.  In the continuum, the densitized and complexified triad $\uo{\tilde{E}}{i}{a}$ together with the Ashtekar connection $\ou{A}{i}{a}$ span $2\times 3\times 3=18$ dimensions over each point on the initial hypersurface. Hence, there is an apparent mismatch between the continuum and the discrete. The additional six complex dimensions are an artefact of working with holonomy flux variables. 
At the discretized level, the holonomy-flux variables $(h_I,E^I)\in SL(2,\C)\times\mathfrak{sl}(2,\C)$ are attached to all four directions $I=1,\dots,4$ of the tetrahedron and they are treated as functionally independent directions of the kinematical phase space. In the continuum, they are not. The tangent space indices $a,b,c\dots$ of the Ashtekar variables $(\uo{\tilde{E}}{i}{a},\ou{A}{i}{a})$ run over a three-dimensional vector space. Hence there seem to be additional spurious directions at the discretized level. To remove the spurious directions, we argued that there are three additional closure constraints missing that would reduce $T^\ast SL(2,\mathbb{C})^4$ to a $18$-dimensional phase space. A proposal for such a constraint was given in \eref{clos2}. The \emph{second result} was to demonstrate that the resulting constraint algebra closes (\hyperref[sec3]{section 4}). It is at this point that the additional closure constraints become crucial. Without the additional constraints, there would be an anomaly. The discretized Gauss, vector and Hamiltonian constraints  do not form a closed algebra among themselves. If we add, however, the dressed closure constraint \eref{clos2}, we end up with an algebra of first-class constraints for each tetrahedron.\medskip

There are many open questions. In our model, there is no matter, the cosmological constant is set to zero and we completely ignored the issue with the reality conditions that reduce the complexified theory to general relativity with a real Lorentzian metric. However, at this stage, the main limitation of our results appears at a more basic level. The Hamiltonian analysis was conducted on a single tetrahedron. The main open problem is to generalize the results to an arbitrary triangulation. A proposal for how to address this problem was given at the end of \hyperref[sec3]{section 4}. For a triangulation with $N$ building blocks, the resulting action \eref{cpldactn} describes a theory of $N$ coupled particles that carve out a trajectory on the superspace of discrete and selfdual geometries on the lattice. Strikingly similar ideas can be found in the group field theory (GFT) approach to quantum gravity  \cite{Freidel:2005qe,Oriti:2014aa,oriti,Carrozza:2020akv}. Group field theory is an approach to quantum gravity where the kinematical wave function for an individual simplicial building block is promoted into a \emph{second-quantized} field operator. In this way, transition amplitudes for simplicial boundary states turn into Feynman amplitudes for an auxiliary quantum field theory on the underlying configuration space, which is a sort-of mini-superspace of geometry (in three spatial dimensions, this is usually $G^4/G$ for gauge groups $G=SU(2)$ or $G=SL(2,\C)$). A large macroscopic geometry is to be modelled from the collective and average behaviour of a large number of GFT quanta (\emph{atoms of geometry}) \cite{Oriti:2016acw}. %
 The results of this paper resonate with this idea. If the analogy is correct, the action \eref{cpldactn} governs a semi-classical sector of a selfdual GFT at fixed particle number with interactions between nearest neighbours. Quantum cosmology would provide a good  test for this scenario. 
 In this way, a link would be established between loop quantum gravity \cite{thiemann, rovelli, alexreview}, loop quantum cosmology \cite{Ashtekar:2006wn,Agullo:2023rqq,Wilson-Ewing:2015lia} and the GFT cosmological sector \cite{Gielen:2016dss,GFTQC,Oriti:2016qtz,deCesare:2016rsf,Wilson-Ewing:2018mrp}. At the present stage, it is unclear whether the model can support this conjecture. Only future research can tell. 

\providecommand{\href}[2]{#2}\begingroup\raggedright\endgroup


\begin{thebibliography}{10}

\bibitem{Loll:2019rdj}
R.~Loll, ``{Quantum Gravity from Causal Dynamical Triangulations: A Review},''
  {\em Class. Quant. Grav.} {\bf 37} (2020), no.~1, 013002,
  \href{http://arXiv.org/abs/1905.08669}{{\tt arXiv:1905.08669}}.

\bibitem{alexreview}
A.~Perez, ``{The Spin-Foam Approach to Quantum Gravity},'' {\em Living Rev.
  Rel.} {\bf 16} (2013), no.~3,
\href{http://arXiv.org/abs/1205.2019}{{\tt arXiv:1205.2019}}.

\bibitem{rovelli}
C.~Rovelli, {\em Quantum Gravity}.
\newblock Cambridge University Press, Cambridge, 2008.

\bibitem{Surya:2019ndm}
S.~Surya, ``{The causal set approach to quantum gravity},'' {\em Living Rev.
  Rel.} {\bf 22} (2019), no.~1, 5, \href{http://arXiv.org/abs/1903.11544}{{\tt
  arXiv:1903.11544}}.

\bibitem{Ambjorn:2022naa}
J.~Ambjorn, ``{Lattice Quantum Gravity: EDT and CDT},''
\newblock 9, 2022.
\newblock \href{http://arXiv.org/abs/2209.06555}{{\tt arXiv:2209.06555}}.

\bibitem{Dona:2022yyn}
P.~Dona, M.~Han, and H.~Liu, ``{Spinfoams and high performance computing},''
  \href{http://arXiv.org/abs/2212.14396}{{\tt arXiv:2212.14396}}.

\bibitem{Dittrich:2014ala}
B.~Dittrich, ``{The continuum limit of loop quantum gravity - a framework for
  solving the theory},'' in {\em Loop Quantum Gravity, The First Thirty Years},
  A.~Abhay and J.~Pullin, eds., vol.~4.
\newblock World Scientific, 2017.
\newblock
\href{http://arXiv.org/abs/1409.1450}{{\tt arXiv:1409.1450}}.
\newblock

\bibitem{Steinhaus:2020lgb}
S.~Steinhaus, ``{Coarse Graining Spin Foam Quantum Gravity\textemdash{}A
  Review},'' {\em Front. in Phys.} {\bf 8} (2020) 295,
  \href{http://arXiv.org/abs/2007.01315}{{\tt arXiv:2007.01315}}.

\bibitem{Asante:2022dnj}
S.~K. Asante, B.~Dittrich, and S.~Steinhaus, ``{Spin foams, Refinement limit
  and Renormalization},'' \href{http://arXiv.org/abs/2211.09578}{{\tt
  arXiv:2211.09578}}.

\bibitem{Benedetti:2007pp}
D.~Benedetti, R.~Loll, and F.~Zamponi, ``{(2+1)-dimensional quantum gravity as
  the continuum limit of Causal Dynamical Triangulations},'' {\em Phys. Rev. D}
  {\bf 76} (2007) 104022, \href{http://arXiv.org/abs/0704.3214}{{\tt
  arXiv:0704.3214}}.

\bibitem{Ambjorn:2020rcn}
J.~Ambjorn, J.~Gizbert-Studnicki, A.~G\"orlich, J.~Jurkiewicz, and R.~Loll,
  ``{Renormalization in quantum theories of geometry},'' {\em Front. in Phys.}
  {\bf 8} (2020) 247, \href{http://arXiv.org/abs/2002.01693}{{\tt
  arXiv:2002.01693}}.

\bibitem{thiemann}
C.~Thiemann, {\em Introduction to Modern Canonical Quantum General Relativity}.
\newblock Cambridge University Press, 2007.

\bibitem{qsd}
T.~Thiemann, ``{Quantum Spin Dynamics (QSD)},'' {\em Class. Quantum Grav.} {\bf
  15} (April, 1998) 839--873, \href{http://arXiv.org/abs/gr-qc/9606089v1}{{\tt
  arXiv:gr-qc/9606089v1}}.

\bibitem{LOSTtheorem}
J.~Lewandowski, A.~Okolow, H.~Sahlmann, and T.~Thiemann, ``{Uniqueness of
  diffeomorphism invariant states on holonomy-flux algebras},'' {\em Commun.
  Math. Phys.} {\bf 267} (2006) 703--733,
\href{http://arXiv.org/abs/gr-qc/0504147}{{\tt arXiv:gr-qc/0504147}}.

\bibitem{Haggard:2023tnj}
H.~M. Haggard, J.~Lewandowski, and H.~Sahlmann, ``{Emergence of Riemannian
  Quantum Geometry},'' \href{http://arXiv.org/abs/2302.02840}{{\tt
  arXiv:2302.02840}}.

\bibitem{outside}
H.~Nicolai, K.~Peeters, and M.~Zamaklar, ``{Loop quantum gravity: an outside
  view},'' {\em Class. Quantum Grav.} {\bf 22} (2005), no.~10, R193--R347,
  \href{http://arXiv.org/abs/hep-th/0501114}{{\tt arXiv:hep-th/0501114}}.

\bibitem{Dittrich:2012qb}
B.~Dittrich, ``{How to construct diffeomorphism symmetry on the lattice},''
  {\em PoS} {\bf QGQGS2011} (2011) 012,
  \href{http://arXiv.org/abs/1201.3840}{{\tt arXiv:1201.3840}}.

\bibitem{Thiemann:2021hpa}
T.~Thiemann and M.~Varadarajan, ``{On Propagation in Loop Quantum Gravity},''
  {\em Universe} {\bf 8} (2022), no.~12, 615,
  \href{http://arXiv.org/abs/2112.03992}{{\tt arXiv:2112.03992}}.

\bibitem{Varadarajan:2022dgg}
M.~Varadarajan, ``{Anomaly free quantum dynamics for Euclidean LQG},''
  \href{http://arXiv.org/abs/2205.10779}{{\tt arXiv:2205.10779}}.

\bibitem{Tomlin:2012qz}
C.~Tomlin and M.~Varadarajan, ``{Towards an Anomaly-Free Quantum Dynamics for a
  Weak Coupling Limit of Euclidean Gravity},'' {\em Phys. Rev. D} {\bf 87}
  (2013), no.~4, 044039, \href{http://arXiv.org/abs/1210.6869}{{\tt
  arXiv:1210.6869}}.

\bibitem{newvariables}
A.~Ashtekar, ``{New Variables for Classical and Quantum Gravity},'' {\em Phys.
  Rev. Lett.} {\bf 57} (1986) 2244--2247.

\bibitem{ashtekar}
A.~Ashtekar, {\em {Lectures on Non-Pertubative Canonical Gravity}}.
\newblock World Scientific, 1991.

\bibitem{Balachandran:1994up}
A.~P. Balachandran, L.~Chandar, and A.~Momen, ``{Edge states in gravity and
  black hole physics},'' {\em Nucl. Phys. B} {\bf 461} (1996) 581--596,
\href{http://arXiv.org/abs/gr-qc/9412019}{{\tt arXiv:gr-qc/9412019}}.

\bibitem{Geiller:2017xad}
M.~Geiller, ``{Edge modes and corner ambiguities in 3d Chern-Simons theory and
  gravity},'' {\em Nucl. Phys. B} {\bf 924} (2017) 312--365,
\href{http://arXiv.org/abs/1703.04748}{{\tt arXiv:1703.04748}}.

\bibitem{Wieland:2017zkf}
W.~Wieland, ``{New boundary variables for classical and quantum gravity on a
  null surface},'' {\em Class. Quantum Grav.} {\bf 34} (2017) 215008,
\href{http://arXiv.org/abs/1704.07391}{{\tt arXiv:1704.07391}}.

\bibitem{Wieland:2017cmf}
W.~Wieland, ``{Fock representation of gravitational boundary modes and the
  discreteness of the area spectrum},'' {\em Ann. Henri Poincar{\'e}} {\bf 18}
  (2017) 3695--3717,
\href{http://arXiv.org/abs/1706.00479}{{\tt arXiv:1706.00479}}.

\bibitem{Donnelly:2016auv}
W.~Donnelly and L.~Freidel, ``{Local subsystems in gauge theory and gravity},''
  {\em JHEP} {\bf 09} (2016) 102, \href{http://arXiv.org/abs/1601.04744}{{\tt
  arXiv:1601.04744}}.

\bibitem{Freidel:2023bnj}
L.~Freidel, M.~Geiller, and W.~Wieland, ``{Corner symmetry and quantum
  geometry},'' \href{http://arXiv.org/abs/2302.12799}{{\tt arXiv:2302.12799}}.

\bibitem{Ashtekar:2020xll}
A.~Ashtekar and M.~Varadarajan, ``{Gravitational Dynamics\textemdash{}A Novel
  Shift in the Hamiltonian Paradigm},'' {\em Universe} {\bf 7} (2021), no.~1,
  13, \href{http://arXiv.org/abs/2012.12094}{{\tt arXiv:2012.12094}}.

\bibitem{Alexander:2022ocp}
S.~Alexander, L.~Freidel, and G.~Herczeg, ``{An Inner Product for 4D Quantum
  Gravity and the Chern-Simons-Kodama State},''
  \href{http://arXiv.org/abs/2212.07446}{{\tt arXiv:2212.07446}}.

\bibitem{Eder:2020okh}
K.~Eder and H.~Sahlmann, ``{Supersymmetric minisuperspace models in self-dual
  loop quantum cosmology},'' {\em JHEP} {\bf 21} (2020) 064,
  \href{http://arXiv.org/abs/2010.15629}{{\tt arXiv:2010.15629}}.

\bibitem{Krasnov:2022mvn}
K.~Krasnov and A.~Shaw, ``{Weyl curvature evolution system for GR},'' {\em
  Class. Quant. Grav.} {\bf 40} (2023), no.~7, 075013,
  \href{http://arXiv.org/abs/2212.12273}{{\tt arXiv:2212.12273}}.

\bibitem{twist3}
L.~Freidel and S.~Speziale, ``{Twisted geometries: A geometric parametrization
  of SU(2) phase space},'' {\em Phys. Rev. D} {\bf 82} (2010), no.~8, 084040,
  \href{http://arXiv.org/abs/1001.2748}{{\tt arXiv:1001.2748}}.

\bibitem{komplexspinors}
W.~Wieland, ``{Twistorial phase space for complex Ashtekar variables},'' {\em
  {Class. Quant. Grav.}} {\bf 29} (2011) 045007,
  \href{http://arXiv.org/abs/1107.5002}{{\tt arXiv:1107.5002}}.

\bibitem{Freidel:2005qe}
L.~Freidel, ``{Group field theory: An Overview},'' {\em Int. J. Theor. Phys.}
  {\bf 44} (2005) 1769--1783, \href{http://arXiv.org/abs/hep-th/0505016}{{\tt
  arXiv:hep-th/0505016}}.

\bibitem{Oriti:2014aa}
D.~Oriti, ``Group Field Theory and Loop Quantum Gravity,'' in {\em Loop Quantum
  Gravity, The First Thirty Years}, A.~Abhay and J.~Pullin, eds., vol.~4.
\newblock World Scientific, 2017.
\newblock \href{http://arXiv.org/abs/1408.7112}{{\tt arXiv:1408.7112}}.

\bibitem{oriti}
D.~Oriti, ``{The group field theory approach to quantum gravity},'' in {\em
  Approaches to Quantum Gravity}.
\newblock Cambridge University Press, Cambridge, 2009.

\bibitem{Carrozza:2020akv}
S.~Carrozza, S.~Gielen, and D.~Oriti, ``{Editorial for the Special Issue
  ''Progress in Group Field Theory and Related Quantum Gravity Formalisms''},''
  {\em Universe} {\bf 6} (2020), no.~1, 19,
  \href{http://arXiv.org/abs/2001.08428}{{\tt arXiv:2001.08428}}.

\bibitem{Oriti:2016acw}
D.~Oriti, ``{The universe as a quantum gravity condensate},'' {\em Comptes
  Rendus Physique} {\bf 18} (2017) 235--245,
\href{http://arXiv.org/abs/1612.09521}{{\tt arXiv:1612.09521}}.

\bibitem{Ashtekar:2006wn}
A.~Ashtekar, T.~Pawlowski, and P.~Singh, ``{Quantum Nature of the Big Bang:
  Improved dynamics},'' {\em Phys. Rev. D} {\bf 74} (2006) 084003,
\href{http://arXiv.org/abs/gr-qc/0607039}{{\tt arXiv:gr-qc/0607039}}.

\bibitem{Agullo:2023rqq}
I.~Agull\'o, A.~Wang, and E.~Wilson-Ewing, ``{Loop quantum cosmology: relation
  between theory and observations},''
  \href{http://arXiv.org/abs/2301.10215}{{\tt arXiv:2301.10215}}.

\bibitem{Wilson-Ewing:2015lia}
E.~Wilson-Ewing, ``{Loop quantum cosmology with self-dual variables},'' {\em
  Phys. Rev. D} {\bf 92} (2015), no.~12, 123536,
  \href{http://arXiv.org/abs/1503.07855}{{\tt arXiv:1503.07855}}.

\bibitem{Gielen:2016dss}
S.~Gielen and L.~Sindoni, ``{Quantum Cosmology from Group Field Theory
  Condensates: a Review},'' {\em SIGMA} {\bf 12} (2016) 082,
  \href{http://arXiv.org/abs/1602.08104}{{\tt arXiv:1602.08104}}.

\bibitem{GFTQC}
E.~Adjei, S.~Gielen, and W.~Wieland, ``{Cosmological evolution as squeezing: a
  toy model for group field cosmology},'' {\em Class. Quant. Grav.} {\bf 35}
  (2018), no.~10, 105016,
\href{http://arXiv.org/abs/1712.07266}{{\tt arXiv:1712.07266}}.

\bibitem{Oriti:2016qtz}
D.~Oriti, L.~Sindoni, and E.~Wilson-Ewing, ``{Emergent Friedmann dynamics with
  a quantum bounce from quantum gravity condensates},'' {\em Class. Quant.
  Grav.} {\bf 33} (2016), no.~22, 224001,
\href{http://arXiv.org/abs/1602.05881}{{\tt arXiv:1602.05881}}.

\bibitem{deCesare:2016rsf}
M.~de~Cesare, A.~G.~A. Pithis, and M.~Sakellariadou, ``{Cosmological
  implications of interacting Group Field Theory models},'' {\em Phys. Rev. D}
  {\bf 94} (2016) 064051, \href{http://arXiv.org/abs/1606.00352}{{\tt
  arXiv:1606.00352}}.

\bibitem{Wilson-Ewing:2018mrp}
E.~Wilson-Ewing, ``{A relational Hamiltonian for group field theory},'' {\em
  Phys. Rev. D} {\bf 99} (2019), no.~8, 086017,
  \href{http://arXiv.org/abs/1810.01259}{{\tt arXiv:1810.01259}}.

\end{thebibliography}

\end{document}